\documentclass[
reprint,
aps,
prapplied,
11pt,
tightenlines,
longbibliography,
showpacs,
nofootinbib,
notitlepage,
superscriptaddress
]{revtex4-2}

\usepackage{amsmath, amssymb, physics, bm}
\usepackage{graphicx}
\usepackage{tabularx}
\usepackage{mathtools}
\usepackage[colorlinks=true, allcolors=black]{hyperref}
\usepackage[capitalize]{cleveref}

\usepackage{subfiles}
\newcommand{\LF}{\mathcal{L}}

\usepackage[shortlabels]{enumitem}
\setlist[enumerate]{leftmargin=20pt, itemsep=-2pt, topsep=5pt}

\begin{document}

\title{Bootstrapping Classical Shadows for Neural Quantum State Tomography}

\author{Wirawat~Kokaew}
\email[These authors contributed equally to this work.]{}
\affiliation{1QB Information Technologies (1QBit), Vancouver, British Columbia, Canada}
\affiliation{Department of Physics \& Astronomy, University of Waterloo, Waterloo, Ontario, Canada}
\affiliation{Perimeter Institute for Theoretical Physics, Waterloo, Ontario, Canada}

\author{Bohdan~Kulchytskyy}
\email[These authors contributed equally to this work.]{}
\affiliation{1QB Information Technologies (1QBit), Vancouver, British Columbia, Canada}

\author{Shunji~Matsuura}
\affiliation{1QB Information Technologies (1QBit), Vancouver, British Columbia, Canada}

\author{Pooya~Ronagh}
\thanks{{\vskip-10pt}{\hskip-9pt}Corresponding author: \href{mailto:pooya.ronagh@1qbit.com}{pooya.ronagh@1qbit.com}\\}
\affiliation{1QB Information Technologies (1QBit), Vancouver, British Columbia, Canada}
\affiliation{Department of Physics \& Astronomy, University of Waterloo, Waterloo, Ontario, Canada}
\affiliation{Perimeter Institute for Theoretical Physics, Waterloo, Ontario, Canada}
\affiliation{Institute for Quantum Computing, University of Waterloo, Waterloo, Ontario, Canada}

\date{\today}

\begin{abstract}
We investigate the advantages of using autoregressive neural quantum states as
ansatze for classical shadow tomography to improve its predictive power.
We introduce a novel estimator for optimizing the cross-entropy loss function
using classical shadows, and a new importance sampling strategy for estimating
the loss gradient during training using stabilizer samples collected from
classical shadows. We show that this loss function can be used to achieve stable
reconstruction of GHZ states using a transformer-based neural network trained
on classical shadow measurements. This loss function also enables the training of
neural quantum states representing purifications of mixed states.
Our results show that the intrinsic capability of autoregressive models
in representing physically well-defined density matrices allows us to overcome
the weakness of Pauli-based classical shadow tomography in predicting both
high-weight observables and nonlinear observables such as the purity of pure
and mixed states.
\end{abstract}

\maketitle
\onecolumngrid

\section{ Introduction } \label{sec:intro}

The accessing of experimentally prepared quantum states is obstructed by the
limited rate at which classical information can be extracted from a quantum
state through repetitive quantum measurements, which is costly. Moreover, in
near-term quantum devices, various sources of physical noise introduce
additional systematic errors, corrupting the measurements. Even when preparing
quantum states fault tolerantly, accessing the logical state's observables of
interest will be susceptible to statistical errors resulting from the finite
number of executions of the fault-tolerant algorithm. For these reasons, it is
imperative to treat the \emph{quantum data} collected from quantum experiments
as scarce, highly valuable information and to maximize the utility of the
available data.

The quantum data must be stored in a classical data structure as a proxy for our
access to information about a given quantum state. In the simplest case, this
data structure is the list of all the measurements performed (e.g., a
tomographically complete set of measurements such as those used in the
classical shadow technique~\cite{Huang2020}).
In principle, such a raw dataset already includes all the information we have
collected from the quantum state. This raises the question as to what the role
of a neural quantum state is when it cannot learn more about the physics of the
state than what it is provided in the training data. However:
\begin{enumerate}[(a)]
\item the amount of memory required to store a tomographically complete set of
 measurements grows exponentially with the size of a quantum system;
\item using this raw data to estimate various observables of the system may
 result in highly erroneous and often unphysical estimations; and
\item every such estimation requires sweeping over the exponentially large list
 at least once.
\end{enumerate}
While traditional quantum state tomography techniques that rely on solving
inverse problems for various (partial or complete) approximations of the
density matrix can overcome the last issue by providing a proxy to the raw
data, they do not resolve the other issues.

Maximum likelihood estimation offers a path to alleviating all these challenges
using memory-efficient parameterized models (i.e., ansatze) and  randomized
subsets of tomographically complete datasets. Model parameters are
variationally optimized in update directions that make the training data the
most likely to be generated by the learned ansatze. Such variational ansatze
can flexibly be used to impose physicality on the reconstructed states. Among
various approaches, tensor networks and neural networks provide flexible
architectures capable of capturing relevant regions within the Hilbert space.
Moreover, when these networks possess an autoregressive property, they can
efficiently provide independent samples for subsequent tasks downstream as
generative models for the learned quantum states.

Recently, Ref.~\cite{wei2023neural} proposed an approach for training a neural
quantum state using classical shadows to combine the advantages of variational
ansatze and classical shadow measurements. The authors introduce the infidelity
between the classical shadow state (introduced in Ref.~\cite{Huang2020}) and
the ansatz as their loss function, since it is provably efficient to estimate
this quantity using Clifford shadow measurements. However, in the case of an
unphysical state like the classical shadow state, the infidelity estimator is
not constrained within the physically valid $[0, 1]$ range. Additionally,
infidelity cannot guarantee bounded errors in approximating many quantities of
interest from the trained model. Finally, the loss does not generalize to
ansatze for mixed states.

Moreover, the method in Ref.~\cite{wei2023neural} faces important practical
challenges in its training protocol. The trainability of the authors' ansatz
relies on an ad hoc initialization step. That is, they pretrain their model
using the cross-entropy loss function on the classical dataset with
measurements only in the computational basis, which works well for quantum
states that are sparse in the computational basis, such as the GHZ state. This
pretraining requires an additional set of measurements beyond the one used in
the classical shadow technique. Furthermore, its success depends on the
informational content of the underlying quantum state in the computational
basis, rendering it non-universal.

In this paper, we introduce a novel estimator for optimizing the cross-entropy
loss function. Our approach capitalizes on the classical shadow state in order
to bootstrap the dataset for construction of a novel unbiased estimators.
Moreover, during training, we advocate for the utilization of stabilizer
samples instead of relying on samples generated from our generative ansatz.
This not only eliminates the necessity for pretraining but also provides a
smoother training gradient. To summarize, the main contributions of our paper
are as follows.
\begin{enumerate}
\item \emph{New loss function.} We provide a novel estimator for the
 cross-entropy loss function using classical shadows. This loss function is
 viable for training mixed states.
\item \emph{New importance sampling strategy.} We introduce
\emph{stabilizer-based} sampling for estimating the overlap between classical
 shadows and a neural quantum state. This overcomes the need for pretraining
 and significantly reduces the variance of gradient estimations during
 training.
\item \emph{Supervision of a physically valid mixed-state autoregressive model
 on measurement data.} We show that our new loss function can be used to train
 a neural network representing a purification of the mixed quantum state from
 classical shadow training data.
\item \emph{Superior prediction of high-weight observables compared to raw
 classical shadow estimations.} We show that the physicality constraints
 natively imposed by the explicit access of autoregressive models to
 conditional probability densities (in the autoregressive expansion) improves
 the accuracy of predictions extracted from classical shadow data.
\end{enumerate}

{\bf Results on pure states.} To showcase the strengths of our ideas, we focus
on the Greenberger--Horne--Zeilinger (GHZ) state that is typically used as a
challenging benchmark for quantum state tomography due to its multi-partite
entanglement. In \cref{sec:results-pure}, we employ Clifford measurements of
pure GHZ states to conduct a comparative study. Our new loss function and
importance sampling strategy show superior performance in learning six- and
eight-qubit GHZ states compared to existing methods~\cite
{wei2023neural, Bennewitz2021}.

{\bf Results on mixed states.} Since in any relevant experimental setup the
states to be studied and characterized are mixed, we use a six-qubit GHZ state
depolarized according to various channel strengths to highlight the
significance of our techniques for experimentalists.
We note that Clifford measurements require deep circuits to implement Clifford
twirling, which is not feasible on noisy quantum computers. To overcome this
challenge, \emph{shallow shadow} protocols have been
\mbox{devised~\cite{ippoliti2023operator, rozon2023optimal, hu2024demonstration,
 hu2022hamiltonian}.} The depth of the Clifford decomposition prescribes an
 upper bound on the weights of the observables that can be estimated well using
 the collected measurements. The generalization power of our natively physical
 ansatz allows us to use the ``shallowest'' possible Clifford tails(i.e., Pauli
 measurements) to achieve satisfactory predictions of high-weight Pauli
 observables as well as nonlinear observables such as purity.
\section{ Methods } \label{sec:methods}

In this section, we review existing methods used for efficiently reconstructing
quantum states. In particular, we consider the recent
development of neural networks and the classical shadow technique for reconstructing valid
quantum states~\cite{wei2023neural, Cha2020, Hu2023}.

\subsection{Neural Quantum States}
\label{sec:neural-q-states}

A promising way to efficiently parameterize a pure quantum state $\ket{\psi}$
of multiple qubits is with a neural network parameterized by a weight vector
$\lambda$ as
\begin{equation}\label{eq:ansatz_pure}
\psi_\lambda(s)=\sqrt{p_\lambda(s)}e^{i \varphi_\lambda(s)},
\end{equation}
where $s \in \mathcal B_n= \{0, 1\}^n$ is a binary vector in the computational
basis. Similarly, the larger state space of density matrices can be
parameterized with a neural network. Among competing approaches, the one based
on the idea of purification~\cite{Torlai2018},
\begin{equation}\label{eq:ansatz_mixed}
\rho_\lambda(s_1, s_2)= \sum_{\bar{s} \in \mathcal B}
\psi^{*}_\lambda(s_1, \bar{s}) \psi_\lambda(s_2, \bar{s}),
\end{equation}
is particularly appealing, as the resulting density matrix is physical. Here,
$\bar{s}$ are auxiliary input variables representing the extended dimensions of
the Hilbert space of the purified state. Therefore, their number $n_{\bar{s}}$
constrains the entropy of $\rho_\lambda$, with
$n_{s}$ being a theoretically proven sufficient upper bound for capturing any density
matrix over the physical input variables~$s$. We treat $n_{\bar{s}}$ as a
hyperparameter on the same footing as other parameters affecting the neural
network's architecture and hence its expressive power.

When the neural network parameters $\lambda$ in the ansatz~\cref{eq:ansatz_pure,eq:ansatz_mixed} are embedded within a generative
model based on an autoregressive architecture, the resulting model can be very
expressive and amenable to an efficient extraction of information about the
underlying wave function as long as the quantity of interest can be cast as an
unbiased estimator over samples from the model. For example, the purity of
$\rho_\lambda$ can be estimated by evaluating the \texttt{Swap} operator
between the physical and auxiliary qubits~\cite
{hasting2010swap, kulchytskyy2019renyi, qucumber2019}.

\subsection{Classical Shadows}
\label{sec:classical-shadows}

The classical shadow formalism provides an efficient method for predicting
various properties using relatively few measurements. A classical shadow is
created by repeatedly performing a simple procedure: at every iteration $i$
a random unitary transformation $\rho\mapsto U_i\rho U_i^\dagger$ is applied, after which
all the qubits in the computational basis are measured. We can efficiently
make a record of the pairs $(U_i, s_i)$ in classical memory. Let
$\rho_i= U^\dagger \ketbra{s_i}{s_i} U$ be the
associated density matrix. The mapping
$\mathcal M$:$~\rho \mapsto \mathbb E[\rho_i]$ is
invertible if the measurement protocol is tomographically complete. The set
$\mathcal S_\rho= \{\mathcal M^{-1} (\rho_i)\}_{i= 1}^N$ is called a
\emph{classical shadow} of $\rho$ of size $N$ and can be used to construct an
unbiased estimator $\hat\rho= \mathbb E[\mathcal M^{-1} (\rho_i)]$ of $\rho$.

The classical shadow $\mathcal S_\rho$ suffices to predict $M$ arbitrary linear
and nonlinear functions $O_i$ of the state $\rho$ up to an additive error $\epsilon$
if it is of size
\mbox{$\Omega(\log(M)~{\rm max}_i~||O_i||^2_{\rm shadow}/\epsilon^2)$.}
The shadow norm depends on the ensemble from which the unitaries $U_i$ are drawn
for creating the classical shadow~\cite{Huang2020} and has efficient bounds,
particularly if the ensemble satisfies a unitary 3-design property. For
example, this is the case if the unitaries are drawn from the Clifford set as in Ref.~\cite
{Huang2020} or using analog evolutions of different lengths of time as in Ref.~\cite{liu2023predicting}.

Two particularly interesting ensembles are the Pauli and Clifford groups. These
ensembles are computationally favourable because of their efficient classical
simulations, according to the Gottesman--Knill theorem. The inverse twirling maps for the $n$-qubit Clifford and Pauli groups take the forms $\mathcal
{M}^{-1}_n\left(X\right) =(2^n + 1) X - \mathbb{I}$ and
$\otimes_{k=1}^n \mathcal M_1^{-1}$, respectively. The shadow norm scales
well for the Clifford ensemble, while the Pauli ensemble is more experimentally
friendly. For the Pauli ensemble, the shadow norm scales exponentially poorly
with respect to the locality of an observable, rendering the procedure
inefficient in estimating high-weight observables. A middle-ground solution
suggested in the literature \cite{ippoliti2023operator, rozon2023optimal,
hu2024demonstration} is to use shallow subsets of the Clifford group.

In this paper, we show that an alternative approach to improving the predictive
power of Pauli ensembles (i.e., the shallowest possible shadows) is using a neural
network (or other parameterized ansatze). Whereas the raw classical shadow is
powerful at predicting many observables efficiently, it is not always sample
efficient due to the unphysical nature of the state it provides. This is
because, while shadow tomography yields a trace-one quantum state, not all
eigenvalues are guaranteed to be positive. However, neural quantum states allow
for natively imposing physicality constraints on a reconstructed state.

Another advantage of the unitary design property of the Clifford ensemble is
that it allows for the efficient prediction of nonlinear functions of a quantum
state. This is particularly useful in estimating entropy-based quantities such
as the R\'enyi entropies, since they involve two copies of the state, and thus
quadratic functions of it. We show that neural networks also improve the
predictive power of such nonlinear quantities when combined with the
purification technique explained in \cref{sec:neural-q-states}.

\subsection{Neural Quantum State Tomography}

\subsubsection{Shadow-Based Loss Functions}

Ref.~\cite{wei2023neural} introduces the idea of using a classical shadow
as a training objective for a neural quantum state. In the paper, the
infidelity between the neural and shadow states from Clifford ensembles,
\begin{equation*}
1 - F_{\lambda} = 1 - \ev{\hat\rho}{\psi_\lambda},
\end{equation*}
is used to derive the loss function
\begin{equation}\label{eq:loss_infidelity}
\LF_{\rm{inf}}^{\rm{C}}(\lambda) = - \sum_{i=1}^{N} p_{\lambda}(\phi_i).
\end{equation}
Here, ``C'' stands for Clifford ensembles, ``inf'' refers to the infidelity, and $p_{\lambda}(\phi_i) = \lvert \braket{\psi_{\lambda}}{\phi_i} \rvert^2 $
is the probability of measuring the snapshot $\phi_i = U_i|s_i\rangle$.

The same loss function can also be modified for the case of Pauli ensembles
as follows. For each Pauli $U_i= U_{i, 1} \otimes \cdots \otimes U_{i, n}$
and any bitstring $b \in \mathcal{B}_n \equiv \{0, 1\}^n$, let $|b|= \sum_{i=1}^n b_i$ be the number
of ones in $b$. Let $U_{i, b}= U_{i, 1}^{b_1} \otimes \cdots \otimes
U_{i, n}^{b_n}$ be the
operator that has an identity factor for each zero index in $b$. In this case,
\begin{align*}
\mathcal M^{-1} \rho_i&= \bigotimes_{k=1}^n (3U_{i,k}^\dagger\ketbra{s_{i,k}}{s_{i,k}} U_{i,k} - I)\\
&= \bigotimes_{k=1}^n (3U_{i,k}^\dagger\ketbra{s_{i,k}}{s_{i,k}} U_{i,k} -
\left[\ketbra{0} + \ketbra{1}\right])\\
&=\sum_{b \in \mathcal B_n} \sum_{c \in \mathcal B_{n-|b|}}
(-1)^{n- |b|} 3^{|b|} U_{i, b}^\dagger \ketbra{s_{i, b, c}} U_{i,b}\\
&=\sum_{b \in \mathcal B_n} \sum_{c \in \mathcal B_{n-|b|}}3^{|b|} \ketbra{\phi_{i, b, c}}.
\end{align*}

Here, $\ket{\phi_{i, b, c}}= (-1)^{\frac{n - |b|}2} U_{i, b}^\dagger
    \ket{s_{i, b, c}}$, where the computational basis state $\ket{s_{i, b, c}}$
    repeats the bit in $s_{i}$ for locations that are hot in the bitstring $b$
    and uses the bit in bitstring $c$ in every other location.
The resulting training loss function is
\begin{equation}\label{eq:loss_infidelity_pauli}
\LF_{\rm{inf}}^{\rm{P}}(\lambda)
= - \sum_{i=1}^{N} \sum_{b \in \mathcal B_n} \sum_{c \in \mathcal B_{n-|b|}} 3^{|b|} p_{\lambda}(\phi_{i, b, c}),
\end{equation}
where ``P'' stands for Pauli ensembles.

We suppress the superscripts ``$\rm{P}$'' and ``$\rm{C}$'' in the rest of this section,
as the discussion that follows applies to both training datasets.
Unfortunately, these loss functions are not generalizable to mixed states. To
circumvent this drawback, we reconsider the Kullback--Leibler (KL) divergence. Our Pauli or
Clifford datasets can be viewed as samples drawn from a target probability
measure $p^{\rm{data}}$:$~\rm{Stab} \to \mathbb R$ on the space of stabilizer
states. For Pauli measurements this is the product of single-qubit stabilizer
states $\mathrm{Stab}_1^{\otimes n}$, and for Clifford measurements it is
$\mathrm{Stab}_n$, the full set of all $n$-qubit stabilizer states. The neural
network (or, more generally, the autoregressive model) representing the quantum
state is a priori a generative model with explicit access to a probability
distribution $p_\lambda$:$~\{0, 1\}^n \to \mathbb R$ on the computational basis
states. However, the efficiency of stabilizer state tableau representation allows
us to extend it to a measure $p_\lambda$:$~\rm{Stab}\to \mathbb R$ on the
stabilizer states. The KL divergence between $p_\lambda$ and $p^{\rm{data}}$ can be written as
\begin{equation*}
    \mathrm{KL}\left( p^{\mathrm{data}} \vert \vert p_{\lambda}\right)
    = \expval{\ln p^{\mathrm{data}}(\phi)}_{\phi \sim p^{\mathrm{data}}}
    - \expval{\ln p_{\lambda}(\phi)}_{\phi \sim p^{\mathrm{data}}}.
\end{equation*}
We note that the entropy of the underlying data distribution is constant with
respect to the variational parameters $\lambda$. Consequently, the loss
function reduces to the cross-entropy term
$-\langle \ln p_{\lambda}(\phi) \rangle_{\phi \sim p^{\mathrm{data}}}$.

Let $\mathcal D \subseteq \rm{Stab}$ be an independent, identically distributed family of samples (allowing
repetitions) drawn from $p^{\mathrm{data}}$, and $\widetilde{\mathcal D}$ be
the set of elements in the dataset $\mathcal D$ (removing repetitions). The cross-entropy
loss function can be approximated by the log-likelihood of $\mathcal D$,
\begin{equation}\label{eq:loss_log_likelihood}
\LF_{\text{ECE}}(\lambda)= - \sum_{\phi \in \mathcal D} \ln p_{\lambda}(\phi),
\end{equation}
which we call the \emph{empirical cross-entropy} (ECE) loss function.

This is consistent with the intuition that the optimal model should treat the
observed samples as the most probable ones. \Cref{eq:loss_log_likelihood} can be viewed in light of the empirical distribution
\begin{equation}\label{eq:empirical_distribution}
p^{\mathrm{emp}}(\phi)
= \frac{1}{|\mathcal D|} \sum_{\phi' \in \mathcal D} \mathbb{I}_{\phi' = \phi} \, ,
\end{equation}
where $\mathbb{I}_{\phi' = \phi}$ is an indicator function. Since
$p^{\mathrm{emp}}$ is an unbiased estimator for $p^{\mathrm{data}}$, the
loss function \cref{eq:loss_log_likelihood} can be rewritten as
\begin{equation*}\label{eq:loss_cross_entropy-empirical}
\LF_{\text{ECE}}(\lambda)
= -\expval{\ln p_{\lambda}(\phi)}_{\phi \sim p^{\mathrm{emp}}}
= -\sum_{\phi \in \widetilde{\mathcal D}} p^{\rm{emp}}(\phi) \ln p_{\lambda}(\phi)\, .
\end{equation*}
We note that the new summation runs over distinct elements of the training data
without repetition. This reformulation helps us to consider alternative
unbiased estimators for the data distribution. Here, we consider an approach based on
the classical shadow state $\hat \rho$. For a stabilizer state $\phi$, we define
its \emph{shadow weight} as
\begin{equation*}
w(\phi) = |\mel{\phi}{\hat \rho}{\phi}|,
\end{equation*}
where the absolute values are taken because $\mel{\phi}{\hat \rho}{\phi}$ can take negative values,
which prevents its being interpreted as a probability distribution. The normalized
shadow weights
\begin{equation}
\label{eq:shadow_weight}
p^{\mathrm{sh}}(\phi) =
\frac{w(\phi)}{\sum_{\phi \in \widetilde{\mathcal D}} w(\phi)}
\end{equation}
can now replace the empirical averaging in the cross-entropy approximation,
justifying the new loss function
\begin{equation}
\label{eq:loss_cross_entropy_shadow}
\LF_{\text{SCE}}(\lambda)
= -\expval{\ln p_{\lambda}(\phi)}_{\phi \sim p^{\mathrm{sh}}}
= -\sum_{\phi \in \widetilde{\mathcal D}} p^{\mathrm{sh}}(\phi) \ln p_{\lambda}(\phi)\,,
\end{equation}
which we call the \emph{shadow-based cross-entropy} (SCE) loss function.
In contrast to the empirical cross-entropy loss function Eq.~(\ref
{eq:loss_log_likelihood}), which asserts equal weights on the contributions of
each stabilizer state $\phi$, the new loss function Eq.~(\ref
{eq:loss_cross_entropy_shadow}) leverages the classical shadow state to
inject further signals useful for the training dynamics.

\subsubsection{Monte Carlo Estimation and Gradients}

Central to the computational tractability of the considered loss functions \cref{eq:loss_infidelity,eq:loss_log_likelihood,eq:loss_cross_entropy_shadow} is the evaluation of the overlap between
a stabilizer state $\phi$ and the neural state $\psi_{\lambda}$:
\begin{equation}
\braket{\psi_{\lambda}}{\phi}
= \sum_{s \in \mathcal B} \psi_{\lambda}^{*}(s) \phi(s).
\label{eq:overlap}
\end{equation}
This exponentially large expansion in the computational basis is intractable.
However, it can be estimated using Monte Carlo sampling in at least two 
ways. We can sample $s \in \mathcal B$ according to
$p_\lambda(s)= \|\psi^*_\lambda(s)\|^2$ to obtain the estimator
\begin{align}
\braket{\psi_{\lambda}}{\phi}
&= \ev{\frac{\phi(s)}{\psi_{\lambda}(s)}}_{s \sim p_{\lambda}}\!.
\label{eq:estimator_lambda}
\end{align}
Alternatively, we can sample $s \in \mathcal B$ with a probability
proportionate to its overlap with the stabilizer state $\phi$,
$p_\phi= \|\phi(s)\|^2$, to obtain the reformulated estimator
\begin{align}
\braket{\psi_{\lambda}}{\phi}
&= \ev{\frac{\psi_{\lambda}^{*}(s)}{\phi^{*}(s)}}_{s \sim p_\phi}\!.
\label{eq:estimator_phi}
\end{align}

Both estimators are efficient to compute, as we can obtain the amplitudes and
samples in polynomial time for both a stabilizer state $\phi$ and an
autoregressive generative model representing $p_{\lambda}$. The question remains as to
which estimator has a better (i.e., lower) variance when used in training the neural
network. The estimator Eq.~(\ref{eq:estimator_lambda}) is employed in
Ref.~\cite{wei2023neural}. This sampling strategy is akin to on-policy reinforcement
learning, wherein the agent $p_{\lambda}$ attempts to optimize its actions $s$
in order to optimize a reward. Hence, just like ordinary policy-gradient methods
based on the REINFORCE technique, training is prone to suffering from a lack
of diversity and therefore highly erroneous gradients.

In this paper, we propose using the estimator
Eq.~(\ref{eq:estimator_phi}) instead, which keeps the training process more
analogous to supervised learning. In this case, the stable statistics collected
from $\phi$ provide a steady target during training. Moreover, unlike the
gradient for Eq.(\ref{eq:estimator_lambda}) (see Ref.~\cite
{wei2023neural}), the gradient for Eq.~(\ref{eq:estimator_phi}) takes the
simpler form
\begin{align*}
\label{eq:estimator_phi}
\nabla_{\lambda}\braket{\psi_{\lambda}}{\phi}
&= \ev{\frac{ \nabla_{\lambda} \psi_{\lambda}^{*}(s)}{\phi^{*}(s)}}_{s \sim p_\phi}.
\end{align*}
%
\section{Results}
\label{sec:results}

\subsection{Scaling and Variance Study}
\label{sec:results-KL}

Prior to studying the training of neural quantum states with our new methods, we
explore the benefits of using shadow weights and stabilizer-based importance
sampling. To this end, we focus on the key training metrics: the informational
content and the statistical properties of the gradient estimator.

\begin{figure}[ht!]
\includegraphics[width=0.8\linewidth]{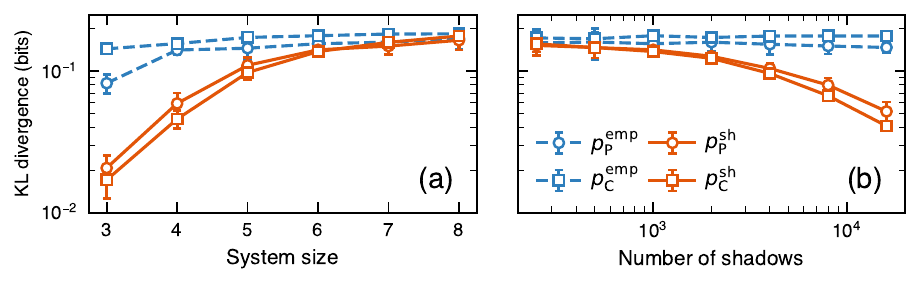}
\caption{
Analysis of the informational content within shadow weights through the KL
divergence ${\rm{KL}}\left( p^{\rm{data}}_{\rm{emp}} \vert \vert p^{\rm
{target}}\right)$. The KL divergence serves as a comparison of the proxy for
the typically unknown data distribution $p^{\rm{data}}_{\rm{emp}} $ (see the
main text for the definition) with the distributions that can be inferred from
the shadow measurements $p^{\rm{emp}}$ based on \cref
{eq:empirical_distribution} and $p^{\rm{sh}}$ specified by  \cref
{eq:shadow_weight}. The KL divergence is examined as a function of (a) the
system size, i.e., the number of qubits, for a dataset of 1000 shadows and
(b) the number of shadows for a six-qubit system. The measurement type is
specified by the marker type: circles and squares for Pauli (indicated by the
subscript ``P'') and Clifford (``C'') measurements, respectively. The target
distribution type is specified by the colours and line styles: blue dashed and
orange solid curves for the empirical distribution and shadow weights,
respectively. Our results are averaged over 32 instances of pure states
produced using random Clifford circuits, with error bars indicating the
standard deviation. }
\label{fig:KL}
\end{figure}

To facilitate our analysis, we consider an empirical proxy for the data
distribution \mbox{$p^{\rm{data}}_{\rm{emp}}(\phi) \propto p^{\rm{data}}
(\phi) \mathbb{I}_{\phi \in \widetilde{\mathcal D}}$,} where $\mathbb{I}_
{\phi \in \widetilde{\mathcal D}}$ is the indicator function that excludes
unobserved stabilizers. This distribution represents the most-accurate
reconstruction of the probability measure $p^{\rm{data}}$ given a dataset
$\mathcal D$, assuming no prior knowledge about $p^{\rm{data}}$. To assess the
informational content in the shadow weights, we consider the KL divergence $
{\rm{KL}}\left( p^{\rm{data}}_{\rm{emp}} \vert \vert p^{\rm
{target}}\right)$ between this typically unknown distribution and $p^{\rm
{target}}$, which denotes either of the two empirically accessible
distributions used for training: $p^{\rm{emp}}$ based on \cref
{eq:empirical_distribution} or $p^{\rm{sh}}$ specified by \cref
{eq:shadow_weight}. Specifically, we study the dependence of ${\rm{KL}}\left
( p^{\rm{data}}_{\rm{emp}} \vert \vert p^{\rm{target}}\right)$ on the size of
the dataset relative to the size of the system by scaling the number of qubits
given a fixed dataset, as shown in Fig.~\ref{fig:KL}(a), and scaling the number
of measurements in the dataset for a fixed number of qubits, as depicted in
Fig.~\ref{fig:KL}(b).

The advantage of shadow weights in terms of estimating $p^{\rm{data}}_{\rm
{emp}} $ is evident for smaller systems, as shown in Fig.~\ref{fig:KL}
(a), though it gradually diminishes as the systems grow larger. In contrast,
Fig.~\ref{fig:KL}(b) indicates that this benefit can be reclaimed by increasing
the number of measurements performed. A crucial factor in interpreting the
plots in Fig.~\ref{fig:KL} is considering the difference in the configuration
space dimensions of $p^{\rm{data}}_{\rm{emp}}$ associated with Pauli and
Clifford measurements. Notably, Pauli measurements can reconstruct $4^n$
distinct stabilizers, whereas Clifford measurements can reconstruct $2^
{0.5 + \mathcal{O}(1)n^2}$ distinct stabilizers~\cite
{aaronson2004improved}. For $n=6$ qubits, the dataset sizes considered in
Fig.~\ref{fig:KL}(b) must be compared to the dimension of the stabilizer space.
For the Pauli distribution, the dataset ranges from 5\% to 500\% of the total
set size of distinct stabilizers. However, for the Clifford distribution, the
maximum number of measurements is less than $0.001 \%$ of the corresponding
stabilizer set size. This significant difference explains the slight scaling
advantage of the $p^{\rm{emp}}$ Pauli measurements over the $p^{\rm
{emp}}$ Clifford measurements, as the former have a higher likelihood of
obtaining repeated stabilizers, while the latter degenerate to a uniform
distribution. For shadow weights, we observe a scaling advantage over empirical
distributions in both regimes. However, unlike in the case of the empirical
distributions, the Clifford distribution shows a slight improvement over the
Pauli distribution. Such an advantage of Clifford shadow weights over Pauli
ones, despite being in a regime with less data, is consistent with the
theoretical expectation of an exponential benefit in predicting high-weight,
low-rank observables (i.e., other stabilizers)~\cite{Huang2020}. Therefore,
given a sufficiently large measurement dataset, training with shadow
weights-based cross-entropy is expected to produce neural quantum states that
better capture the underlying quantum state.

\begin{figure}[h!]
\includegraphics[width=0.8\linewidth]{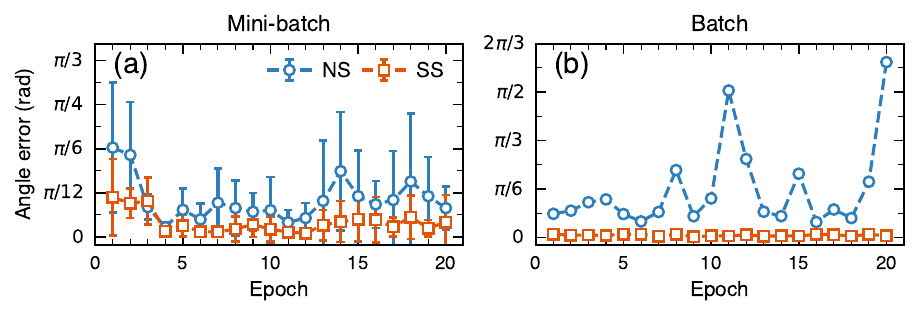}
\caption{
Discrepancy in angle for the Monte Carlo-estimated gradients compared to the
exact gradient throughout the training process. The dataset consists of 1000
Clifford measurements performed on a three-qubit GHZ state. Estimation using
neural state-based sampling (NS) and stabilizer-based sampling (SS) are
depicted in blue and orange, respectively. Each estimation uses 500 Monte Carlo
samples. The gradient errors are evaluated over \mbox{(a) mini-batches} and
(b) a full batch. In the case of the mini-batch gradient error, the error bars
correspond to the standard deviation in the error over all mini-batches.}
\label{fig:sampling}
\end{figure}

We now evaluate the advantages of the proposed stabilizer-based method of \cref
{eq:estimator_phi}, over the neural state-based method of \cref
{eq:estimator_lambda}, for Monte Carlo estimation of the overlap, which is used
to compute the training gradient. For benchmarking purposes, we analyze the
errors in the gradient estimates relative to the true gradient, obtained
through an exhaustive evaluation of the inner product of~\cref
{eq:overlap}. Specifically, we focus on the error in the gradient direction and
distinguish between errors calculated over a mini-batch and a full batch of
measurements. Figure~\ref{fig:sampling} illustrates the evolution of these
errors as the neural quantum state is trained using the true gradient.
Figure~\ref{fig:sampling}(a) shows that both estimators struggle most with
capturing the correct gradient direction during the early stages of training,
where the neural quantum state is likely the most different from the target
state. However, after several epochs, as the neural state becomes more similar
to the target, the stabilizer-based estimation significantly improves, while
the neural state estimation remains relatively high and inconsistent.
Throughout the entire training process, the stabilizer-based estimator
consistently provides estimates with much lower variance, which is essential
for training stability. Additionally, our proposed estimator is consistently
more accurate than the neural-based estimator. Being an unbiased estimator, the
stochastic error due to mini-batching is expected to be cancelled out when
averaged over the mini-batches and thus can be lowered using smaller
learning rates, albeit at the cost of slower training dynamics. Figure~\ref
{fig:sampling}(b) shows whether this expected behaviour holds for the given
estimators. There is a clear difference between them: the stabilizer-based
estimator functions as an unbiased estimator, reducing gradient errors by
averaging over mini-batches. In contrast, the neural state-based estimator
exhibits a systematic bias throughout the entire training duration, resulting
in an extreme accumulation of errors from the true gradient. To eliminate this
finite-sample bias, a significant increase in the samples generated by the
neural state would be necessary. Therefore, given a fixed computational
resource budget, our benchmarking results demonstrate that the stabilizer-based
gradient estimation has a significant advantage over the neural state-based
gradient estimation, highlighting its potential for optimization.

\subsection{Training Performance on Pure GHZ States}
\label{sec:results-pure}

In this section, we focus on the reconstruction of pure GHZ states from Pauli
and Clifford measurements as a benchmark for the training enhancements we have
proposed. In particular, we use the shadow-based cross-entropy ($\LF_{\text
{SCE}}$), the empirical cross-entropy ($\LF_{\text{ECE}}$), and the infidelity
($\LF_{\text{inf}}$) as the loss functions. In addition, we employ the
estimators Eqs.~(\ref{eq:estimator_lambda}) and~(\ref{eq:estimator_phi}) in
evaluating the overlap between the neural network and stabilizer states. The
superscripts ``NS'' (for neural state-based samples) and ``SS''
(for stabilizer-based samples) appear with the corresponding loss function
$\LF$ to distinguish between the two estimators.

\begin{figure}[h!]
\includegraphics[width=1.0\linewidth]{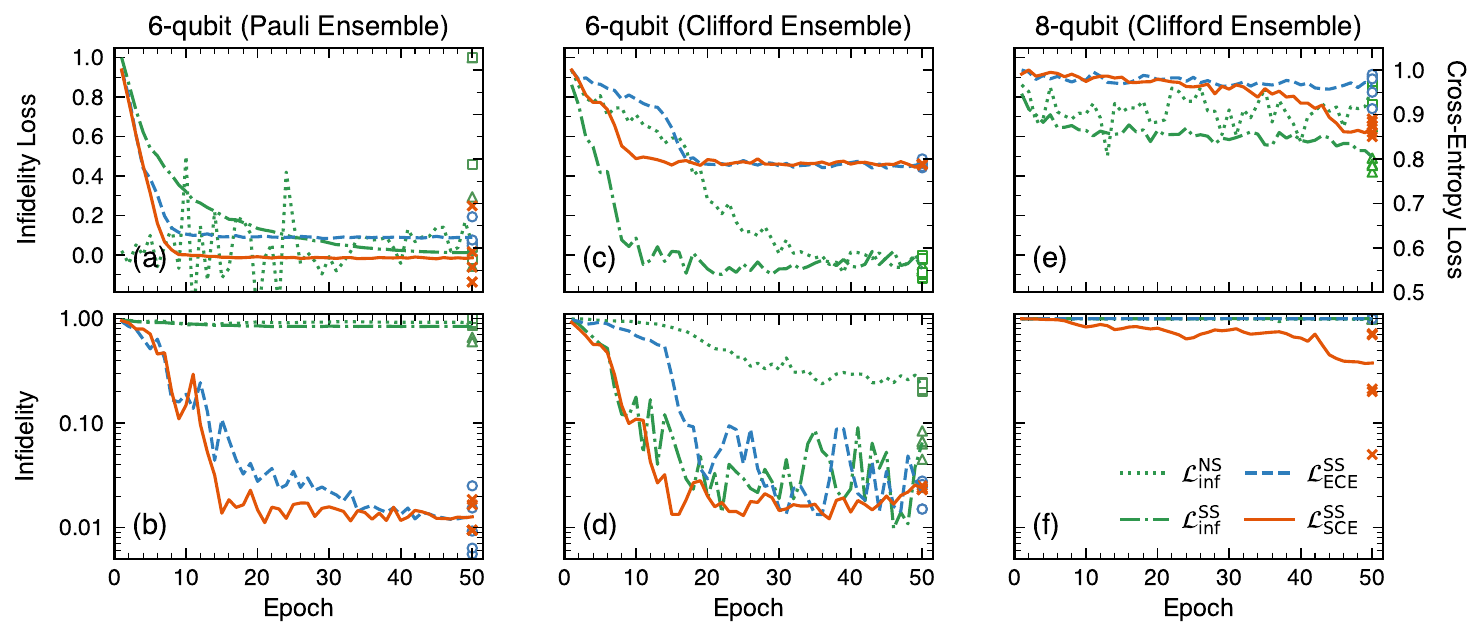}
\caption{Benchmarking results of the performance of four methods in
reconstructing both six- and eight-qubit GHZ states. We conduct our
experiments using 1000 classical shadows based on Pauli or Clifford
measurements (as specified in the titles above the plots), averaging the
results over five training trials on the same dataset trained over 50 epochs.
Each estimate of the overlap between a neural quantum state and a stabilizer
(see \cref{eq:overlap}) is based on 500 newly generated samples. The final
states for each loss function are represented by markers. Triangles and square markers represent the samples from the last epoch for
the stabilizer- and neural state-based infidelity loss functions, respectively. Circles and ``x'' markers represent the samples from the last epoch for
the empirical and shadow-based cross-entropy loss functions, respectively. The upper row displays the infidelity loss (left y-axis) and the cross-entropy loss (right y-axis) for the loss functions while the lower row shows the
associated infidelity with respect to the true state as a measure of the
generalization power of the neural quantum state. In order to facilitate comparison with the infidelity loss, the range of the cross-entropy loss is normalized not to exceed 1. }
\label{fig:benckmark}
\end{figure}

We illustrate the trainability and the generalization power of these training
methods in Fig.~\ref{fig:benckmark} via the optimization progress
curves of loss functions evaluated on the training set (upper row) and
infidelity with respect to the true states (lower row), over 50 epochs. For Pauli
measurements, plotted in Fig.~\ref{fig:benckmark}(a)~and~(b), the superiority
of cross-entropy-based training over the infidelity-based approach is apparent.
The latter struggles to achieve significant progress in approximating the
underlying quantum state. Moreover, the benefits of the cross-entropy-based
loss function become even more pronounced when considering computational costs.
Both loss functions rely on computing the overlap between the neural quantum
state and stabilizers (see \cref{eq:overlap}). For the infidelity loss
function, each Pauli measurement leads to a proliferation of stabilizers that
grows super-exponentially with the number of qubits (see \cref
{eq:loss_infidelity_pauli}). In contrast, this prohibitive scaling is entirely
absent for both cross-entropy-based loss functions, as each Pauli measurement
contributes a single stabilizer. In a previous work~\cite{wei2023neural}, we attempted to address the prohibitive
scaling of the infidelity-based approach through Monte Carlo sampling of
stabilizers from a classical shadow state. However, this strategy led to
unstable training. Additionally, we note that stabilizer-based training
improves in-training performance across all loss functions (the results for
neural-state-based cross-entropy loss functions have been omitted for visual
clarity). Moreover, these improvements during training translate to enhanced
generalization only for cross-entropy loss functions. In fact, for the
cross-entropy loss function, stabilizer-based training proves to be crucial.

We now discuss the Clifford measurements, beginning with the benchmarking
performed on the six-qubit GHZ state shown in Fig.~\ref{fig:benckmark}(c)~and~(d). Our previous observations regarding the advantages of stabilizer-based
sampling generalize to this scenario and, in fact, are applicable to the
infidelity loss function as well. Indeed, the loss function $\LF_{\text
{inf}}^\text{SS}$ leads to an improved generalization with the final infidelity
of roughly an order of magnitude compared to the loss function $\LF_
{\text{inf}}^\text{NS}$, as shown in Fig.~\ref{fig:benckmark}(d). Interestingly, this improvement in generalization is achieved despite
their corresponding loss functions yielding similar values at the end of
training. Further examination of Fig.~\ref{fig:benckmark}(c)~and~(d) reveals
that our loss function $\LF_{\text{ECE}}^{\text{SS}}$ greatly outperforms the
previously proposed shadow-based loss function $\LF_{\text{inf}}^\text
{NS}$~\cite{wei2023neural} in both the convergence rate and generalization.
In addition, a comparison between the loss functions $\LF_{\text{SCE}}^{\text
{SS}}$ and $\LF_{\text{ECE}}^{\text{SS}}$ reveals that our cross-entropy
estimator leads to a superior convergence rate and lower infidelity,
illustrating the benefits of shadow weights extracted from the classical shadow
state (see Eq.~(\ref{eq:shadow_weight})). Overall, both loss functions $\LF_
{\text{SCE}}^{\text{SS}}$ and $\LF_{\text{inf}}^\text{SS}$  demonstrate the
best convergence rates. The loss functions $\LF_{\text{SCE}}^{\text{SS}}$ and
$\LF_{\text{ECE}}^{\text{SS}}$ exhibit the greatest generalization, approaching
an infidelity of $0.01$, while our estimator shows the greatest stability in reaching
states in proximity to the true states. An even more drastic difference in
performance is observed for the eight-qubit GHZ state, shown in Fig.~\ref
{fig:benckmark}(e)~and~(f). In this case, the loss functions $\LF_{\text
{inf}}^\text{NS}$, $\LF_{\text{inf}}^\text{SS}$, and $\LF_{\text{ECE}}^{\text
{SS}}$ struggle to escape local minima. While all of the loss functions result
in infidelities that remain stuck at $0.99$, only the loss function $\LF_{\text
{SCE}}^{\text{SS}}$ stands out, achieving an infidelity as low as $0.05$. To
summarize, our results indicate that our proposed method provides the best
training objective for neural networks trained to approximate an underlying
quantum state from limited measurements.

\subsection{Applications on Mixed GHZ States}
\label{sec:results-mixed}

In this section, we describe our application of the methods introduced earlier, namely, the
shadow-based cross-entropy loss function with stabilizer-based sampling, to
mixed quantum states generated by noisy circuits. Specifically, we consider a
circuit that prepares a six-qubit GHZ state beginning from the all-zeroes state
$\ket{0}^{\otimes 6}$. The circuit is composed of a Hadamard gate applied to
the first qubit, followed by a series of CNOT gates applied to the consecutive
pairs of qubits. For the noise model, we assume two-qubit depolarization noise
on the CNOT gates. The depolarizing noise channel $\rho \mapsto (1-\frac{16}
{15}p) \rho + \frac{16}{15}p \mathbb{I}/4$ is parameterized via a noise
strength parameter $p$. Motivated by the experimental relevance of our
technique, we consider Pauli measurement datasets, as those are less noisy
than Clifford measurements. As the ansatz for the density matrix, we employ the
purification ansatz Eq.~(\ref{eq:ansatz_mixed}) with a transformer neural
network used to parameterize $\psi_{\lambda}(s, \bar{s})$. This approach allows
us to impose physical constraints on the ansatz.

\subsubsection{Pauli Observables}
Although classical shadows based on Pauli measurements have been demonstrated to
predict local observables with high accuracy, their predictive accuracy worsens
in the case of many-body observables. In this section, we demonstrate the
benefits of the neural network approach over classical shadows in predicting
Pauli observables. To do so, a set of 5000 random Pauli strings is
generated by sampling each single-qubit Pauli operator from a uniform distribution.

\begin{figure}[h!]
\includegraphics[width=0.8\linewidth]{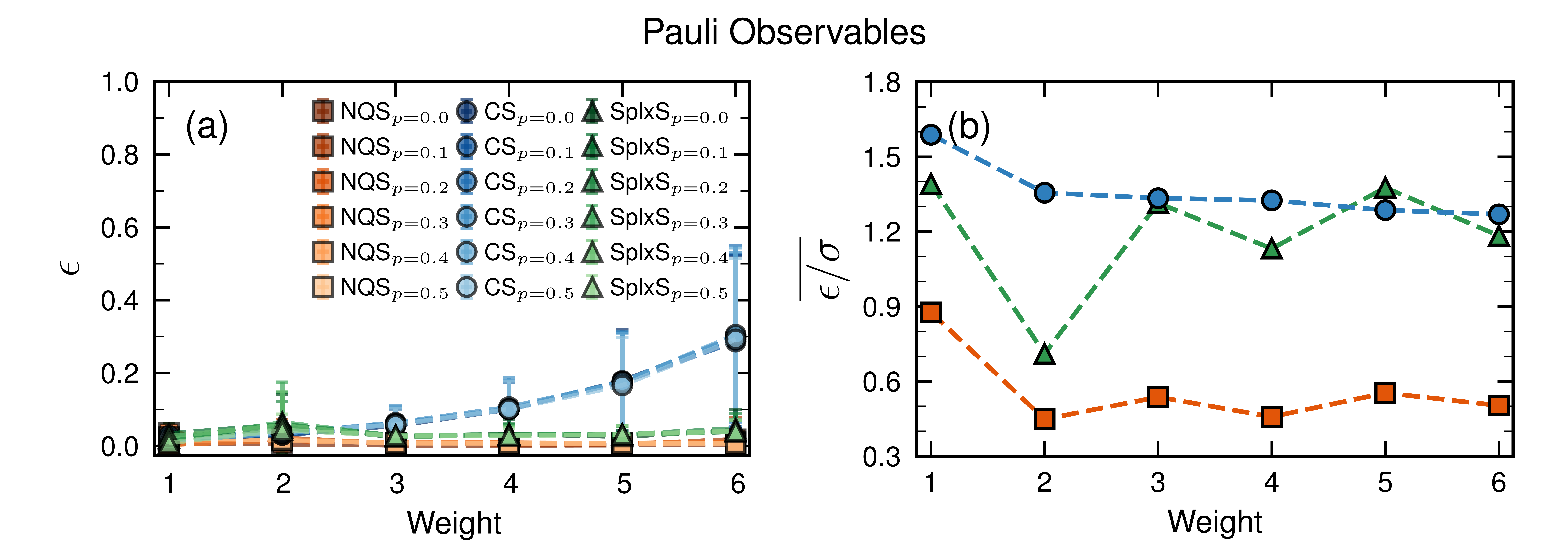}
\caption{Predictive capabilities based on neural quantum states, classical
 shadows, and the simplex projection of the latter, for noisy observables at
 noise levels $p$ ranging from $0.0$ to $0.5$ sampled from 5000 random
 Pauli measurements. (a) Absolute error, $\epsilon$, in predicting random noisy
 observables with standard deviations represented by error bars. (b) Accuracy,
 defined as the average of the absolute error over the standard deviation, $\overline
 {\epsilon/\sigma}$. The results are differentiated by colour, with
 neural quantum states shown in orange, classical shadows in blue, and the simplex 
 projections of classical shadows in green.}
\label{fig:random_pauli-zeronoise}
\end{figure}

The absolute error in predicting Pauli observables from neural networks and
classical shadows is shown in \cref{fig:random_pauli-zeronoise}(a). As
expected, the precision and accuracy of estimates extracted from the classical
shadow technique become larger as the Pauli weight increases. This poor
performance in predicting high weight observables is alleviated by projecting a
classical shadow state to a physical state on the probability simplex
(see \cref{app:simplex}). Despite this adjustment, the neural state ansatz
consistently demonstrates superior performance across all noise levels. This
observation is further supported when focusing solely on prediction accuracy,
as highlighted in \cref{fig:random_pauli-zeronoise}(b), where we present the
same dataset, now normalized by variance and averaged across all noise
levels.

\subsubsection{Purity and Trace Distance}
\begin{figure}[h!]
\includegraphics[width=1\linewidth]{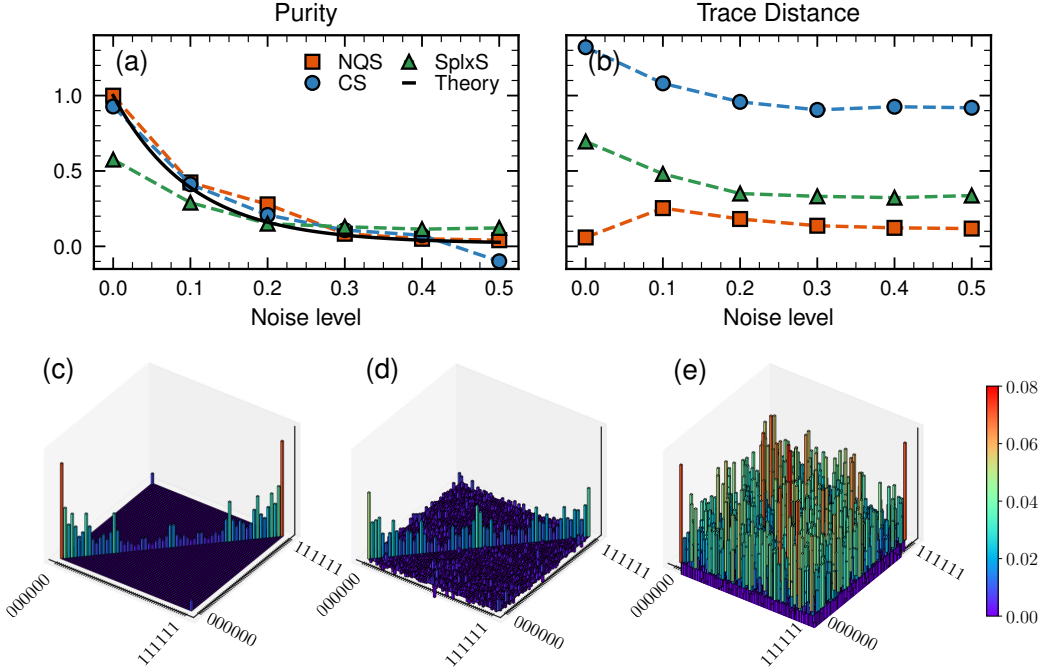}
\caption{Performance comparison of reconstructed six-qubit GHZ mixed states
 using classical shadows and neural quantum states collected from 5000
 random Pauli measurements. (a) Theoretical values of purity for the ideal
 state (shown using a solid curve), along with the reconstructed states from
 neural quantum states (shown in orange), classical shadows (shown in blue), 
 and the simplex projections of classical shadows (shown in green).
 (b) Trace distance with respect to the ideal state. (c)--(e) Values of real
 components of the density matrix elements for the noise level $p=0.5$,
 comparing the ideal case (c), the neural quantum state (d), and the classical
 shadow state (e). The colour bar indicates the magnitude of each component of
 the density matrix.}
\label{fig:dm}
\end{figure}

Continuing our analysis, we broaden our scope to include assessments of purity
and trace distance, shown in Fig.~\ref{fig:dm}(a) and~(b), respectively. In
contrast to the behaviour observed in the case of Pauli observables, projecting the
classical shadow state onto the probability simplex leads to a degradation in
purity estimation, as shown in Fig.~\ref{fig:dm}(a). However, both neural quantum
states and classical shadow states are able to accurately  track theoretically
predicted results. The distinguishing feature of the physically constrained
neural quantum state lies in its ability to consistently provide physical
estimates, whereas the classical shadow state's estimate falls below zero for
the least pure target state we consider. As expected, the observed trends of
the trace distance shown in Fig.~\ref{fig:dm}(b) align with the results of the
Pauli observables: the simplex projection demonstrates successful improvement
over the classical shadow state, while the neural quantum state exhibits the
lowest error for all noise levels. 

Interestingly, we observe relatively weaker performance of the neural quantum
states in the low noise regime, specifically $p=0.1$ and $0.2$. We investigate
this behaviour further by producing visualizations of density matrices, as exemplified in
Fig.~\ref{fig:dm}(c), (d), and~(e) for $p=0.5$. For such high noise levels, all
non-zero components in the theoretical density matrix are substantial. In this
case, the neural quantum state can successfully capture the full structure of
the theoretical density matrix, whereas the classical shadow state is extremely
noisy.  However, at lower noise regimes, most of the diagonal elements of the
density matrix decrease, and only the corner components remain large. The
neural quantum state seems to struggle with capturing signals from very small
components, possibly due to the Monte Carlo sampling noise present during training,
leading to the relative weakening of its performance. We defer a detailed
investigation of this phenomenon and the potential enhancements it could bring
to neural quantum state tomography in this regime to future studies.

\section{Conclusion}
\label{sec:conclusion}

In this paper, we have introduced two new concepts for neural quantum state
tomography: \mbox{(a) the} shadow-based cross-entropy loss function; and
(b) stabilizer-based importance sampling for the estimation of quantum state
overlaps. Our benchmarking using Pauli and Clifford measurements collected from
pure GHZ states demonstrates a clear improvement in the trainability and the
generalization power of the resulting neural quantum states. In particular, we
demonstrate the superiority of these neural quantum states in comparison to the
conventional neural quantum states trained using log-likelihood optimization
and the recently introduced neural quantum states trained using infidelity
loss, both of which rely on neural-state-based importance sampling. These
advancements enable a significant reduction in both the classical and quantum
resources necessary for neural quantum state tomography.

To highlight the significance of our techniques in experimental applications, we
employ a physically constrained neural quantum state to reconstruct mixed
quantum states from Pauli measurements. In comparison to the classical shadow
state constructed from the same dataset, our neural quantum state is much more
successful at predicting high-weight and nonlinear observables of both pure and
mixed states.

With these promising results, we expect shadow-based neural quantum state
tomography to be an invaluable tool for future experiments on near-term quantum
devices. Its potential applications span a wide spectrum, from device
characterization to quantum simulations in materials science, chemistry, and
strong interactions physics. Moving forward, the next steps in the development
and exploration of our method will involve scaling it up to accommodate larger
systems and validating its utility to a more diverse set of quantum states,
specially those prepared experimentally in the near term. One interesting
avenue for future research is the characterization of quantum states prepared
using analog quantum simulations, wherein the constraints imposed by limited-control electronics present a challenge for quantum state tomography~\cite{hu2022hamiltonian}.

\appendix
\section{The Transformer Neural Quantum State and Details of Numerical Experiments}
\label{app:transformer}

We base our neural quantum states on the transformer architecture from
Ref.~\cite{bennewitz2021} (for details on its implementation, refer to Section I of the
supplementary material of that work). More specifically, a
quantum state is represented using the transformer model parameterized by the
neural network weights $\lambda$ as specified in \cref{eq:ansatz_pure} or \cref
{eq:ansatz_mixed}, depending on whether the target quantum state is a pure state or a
mixed state, respectively. The details of the transformer's architecture,
hyperparameters, and datasets are presented in Table~\ref
{table1}. PyTorch~\cite{Paszke2019} is used for the implementation and training
of the transformer. The Stim stabilizer-circuit simulator~\cite{Gidney2021} is
used for dataset generation and classical shadow state reconstruction. For
training, we use the Adam
optimizer~\cite{kingma2014adam} for stochastic-gradient descent. A cosine annealing schedule is applied to
systematically adjust the learning rate throughout training epochs~\cite
{Loshchilov2016COSINE}. Finally, early stopping is used to prevent
over-fitting.

\begin{table}[ht!]
\begin{center}
\begin{tabularx}{1.0\textwidth} { | >{\centering\arraybackslash}X | >
 {\raggedright\arraybackslash}X | >{\raggedright\arraybackslash}X | >
 {\raggedright\arraybackslash}X | }
 \hline
&& \cref{fig:benckmark} & \cref{fig:random_pauli-zeronoise,fig:dm}\\
 \hline
&Layers & 2  & 2 \\
&Internal dimensions & 8  & 8  \\ Neural quantum state&Heads & 4  & 4  \\
&Trainable parameters & 858  & 954  \\
&Ansatz  & \cref{eq:ansatz_pure}  &  \cref{eq:ansatz_mixed}  with 6 ancilla
 qubits  \\
\hline
&Epochs & 50  & 100  \\ Hyperparameters&Initial learning rate & 0.01  & 0.01  \\
&Mini-batch size & 100  & 20  \\
&Monte Carlo samples & 500  & 500  \\
\hline Dataset &Training  & 1000 classical shadows & 3750 classical shadows\\
&Validation & --  & 1250 classical shadows\\
\hline
\end{tabularx}
\caption{Numerical experiment settings for \cref
 {fig:benckmark,fig:random_pauli-zeronoise,fig:dm}.}
\label{table1}
\end{center}
\end{table}

\section{Probability Simplex Projection}
\label{app:simplex}

Unlike the density matrix of a physical quantum state, the classical shadow
state is not necessarily positive semidefinite. One approach for obtaining a
physical quantum state from a classical shadow state is through a projection
onto the probability simplex. Formally, the eigenvalues ${\lambda}\in\mathbb
{R}^{2^n}$ of the classical shadow state are projected onto the probability
simplex $\Delta^{2^n}:=\{x = (x_1,\dots,x_{2^n})^T|~x_i\geq 0
~\wedge~\sum_{i=1}^{2^n} x_i=1\}$, while the eigenvectors are left unchanged.
The projected  eigenvalues, $\lambda^{*}$, are obtained via a minimization of
the Euclidean distance:
\begin{equation*}
    \lambda^{*}={\arg\min_{x\in\Delta^{2^n}}||x-\lambda||},
\end{equation*}
where $||\cdot||$ denotes the Euclidean norm. We use the
 algorithm presented in Ref.~\cite{Chen2011Simplex} to perform this
 minimization.


\section*{Acknowledgements}

We thank our editor, Marko Bucyk, for his careful review and editing of the manuscript.
The authors acknowledge Bill~Coish and Christine~Muschik for useful discussions.
W.~K.~acknowledges the support of Mitacs and a scholarship through the Perimeter
Scholars International program.
P.~R.~acknowledges the financial support of Mike and Ophelia Lazaridis,
Innovation, Science and Economic Development Canada (ISED), and the Perimeter
Institute for Theoretical Physics.
Research at the Perimeter Institute is supported in part by the Government of
Canada through ISED and by the Province of Ontario through the Ministry of
Colleges and Universities.

\bibliography{references}

\begin{thebibliography}{20}%
\makeatletter
\providecommand \@ifxundefined [1]{%
 \@ifx{#1\undefined}
}%
\providecommand \@ifnum [1]{%
 \ifnum #1\expandafter \@firstoftwo
 \else \expandafter \@secondoftwo
 \fi
}%
\providecommand \@ifx [1]{%
 \ifx #1\expandafter \@firstoftwo
 \else \expandafter \@secondoftwo
 \fi
}%
\providecommand \natexlab [1]{#1}%
\providecommand \enquote  [1]{``#1''}%
\providecommand \bibnamefont  [1]{#1}%
\providecommand \bibfnamefont [1]{#1}%
\providecommand \citenamefont [1]{#1}%
\providecommand \href@noop [0]{\@secondoftwo}%
\providecommand \href [0]{\begingroup \@sanitize@url \@href}%
\providecommand \@href[1]{\@@startlink{#1}\@@href}%
\providecommand \@@href[1]{\endgroup#1\@@endlink}%
\providecommand \@sanitize@url [0]{\catcode `\\12\catcode `\$12\catcode
  `\&12\catcode `\#12\catcode `\^12\catcode `\_12\catcode `\%12\relax}%
\providecommand \@@startlink[1]{}%
\providecommand \@@endlink[0]{}%
\providecommand \url  [0]{\begingroup\@sanitize@url \@url }%
\providecommand \@url [1]{\endgroup\@href {#1}{\urlprefix }}%
\providecommand \urlprefix  [0]{URL }%
\providecommand \Eprint [0]{\href }%
\providecommand \doibase [0]{https://doi.org/}%
\providecommand \selectlanguage [0]{\@gobble}%
\providecommand \bibinfo  [0]{\@secondoftwo}%
\providecommand \bibfield  [0]{\@secondoftwo}%
\providecommand \translation [1]{[#1]}%
\providecommand \BibitemOpen [0]{}%
\providecommand \bibitemStop [0]{}%
\providecommand \bibitemNoStop [0]{.\EOS\space}%
\providecommand \EOS [0]{\spacefactor3000\relax}%
\providecommand \BibitemShut  [1]{\csname bibitem#1\endcsname}%
\let\auto@bib@innerbib\@empty
\bibitem [{\citenamefont {Huang}\ \emph {et~al.}(2020)\citenamefont {Huang},
  \citenamefont {Kueng},\ and\ \citenamefont {Preskill}}]{Huang2020}%
  \BibitemOpen
  \bibfield  {author} {\bibinfo {author} {\bibfnamefont {H.-Y.}\ \bibnamefont
  {Huang}}, \bibinfo {author} {\bibfnamefont {R.}~\bibnamefont {Kueng}},\ and\
  \bibinfo {author} {\bibfnamefont {J.}~\bibnamefont {Preskill}},\ }\bibfield
  {title} {\bibinfo {title} {Predicting many properties of a quantum system
  from very few measurements},\ }\href
  {https://doi.org/10.1038/s41567-020-0932-7} {\bibfield  {journal} {\bibinfo
  {journal} {Nat. Phys.}\ }\textbf {\bibinfo {volume} {16}},\ \bibinfo {pages}
  {1050–1057} (\bibinfo {year} {2020})}\BibitemShut {NoStop}%
\bibitem [{\citenamefont {{Wei}}\ \emph {et~al.}(2023)\citenamefont {{Wei}},
  \citenamefont {{Coish}}, \citenamefont {{Ronagh}},\ and\ \citenamefont
  {{Muschik}}}]{wei2023neural}%
  \BibitemOpen
  \bibfield  {author} {\bibinfo {author} {\bibfnamefont {V.}~\bibnamefont
  {{Wei}}}, \bibinfo {author} {\bibfnamefont {W.~A.}\ \bibnamefont {{Coish}}},
  \bibinfo {author} {\bibfnamefont {P.}~\bibnamefont {{Ronagh}}},\ and\
  \bibinfo {author} {\bibfnamefont {C.~A.}\ \bibnamefont {{Muschik}}},\
  }\bibfield  {title} {\bibinfo {title} {{Neural-Shadow Quantum State
  Tomography}},\ }\href {https://arxiv.org/abs/2305.01078} {\bibfield
  {journal} {\bibinfo  {journal} {arXiv preprint arXiv:2305.01078}\ } (\bibinfo
  {year} {2023})}\BibitemShut {NoStop}%
\bibitem [{\citenamefont {Bennewitz}\ \emph {et~al.}(2021)\citenamefont
  {Bennewitz}, \citenamefont {Hopfmueller}, \citenamefont {Kulchytskyy},
  \citenamefont {Carrasquilla},\ and\ \citenamefont {Ronagh}}]{Bennewitz2021}%
  \BibitemOpen
  \bibfield  {author} {\bibinfo {author} {\bibfnamefont {E.~R.}\ \bibnamefont
  {Bennewitz}}, \bibinfo {author} {\bibfnamefont {F.}~\bibnamefont
  {Hopfmueller}}, \bibinfo {author} {\bibfnamefont {B.}~\bibnamefont
  {Kulchytskyy}}, \bibinfo {author} {\bibfnamefont {J.}~\bibnamefont
  {Carrasquilla}},\ and\ \bibinfo {author} {\bibfnamefont {P.}~\bibnamefont
  {Ronagh}},\ }\bibfield  {title} {\bibinfo {title} {{Neural Error Mitigation
  of Near-Term Quantum Simulations}},\ }\href
  {https://doi.org/10.1038/s42256-022-00509-0} {\bibfield  {journal} {\bibinfo
  {journal} {Nat. Mach. Intell.}\ }\textbf {\bibinfo {volume} {4}},\ \bibinfo
  {pages} {618–624} (\bibinfo {year} {2021})}\BibitemShut {NoStop}%
\bibitem [{\citenamefont {Ippoliti}\ \emph {et~al.}(2023)\citenamefont
  {Ippoliti}, \citenamefont {Li}, \citenamefont {Rakovszky},\ and\
  \citenamefont {Khemani}}]{ippoliti2023operator}%
  \BibitemOpen
  \bibfield  {author} {\bibinfo {author} {\bibfnamefont {M.}~\bibnamefont
  {Ippoliti}}, \bibinfo {author} {\bibfnamefont {Y.}~\bibnamefont {Li}},
  \bibinfo {author} {\bibfnamefont {T.}~\bibnamefont {Rakovszky}},\ and\
  \bibinfo {author} {\bibfnamefont {V.}~\bibnamefont {Khemani}},\ }\bibfield
  {title} {\bibinfo {title} {{Operator Relaxation and the Optimal Depth of
  Classical Shadows}},\ }\href {https://doi.org/10.1103/PhysRevLett.130.230403}
  {\bibfield  {journal} {\bibinfo  {journal} {Phys. Rev. Lett.}\ }\textbf
  {\bibinfo {volume} {130}},\ \bibinfo {pages} {230403} (\bibinfo {year}
  {2023})}\BibitemShut {NoStop}%
\bibitem [{\citenamefont {Rozon}\ \emph {et~al.}(2023)\citenamefont {Rozon},
  \citenamefont {Bao},\ and\ \citenamefont {Agarwal}}]{rozon2023optimal}%
  \BibitemOpen
  \bibfield  {author} {\bibinfo {author} {\bibfnamefont {P.-G.}\ \bibnamefont
  {Rozon}}, \bibinfo {author} {\bibfnamefont {N.}~\bibnamefont {Bao}},\ and\
  \bibinfo {author} {\bibfnamefont {K.}~\bibnamefont {Agarwal}},\ }\bibfield
  {title} {\bibinfo {title} {Optimal twirling depths for shadow tomography in
  the presence of noise},\ }\href {https://doi.org/10.48550/arXiv.2311.10137}
  {\bibfield  {journal} {\bibinfo  {journal} {arXiv preprint arXiv:2311.10137}\
  } (\bibinfo {year} {2023})}\BibitemShut {NoStop}%
\bibitem [{\citenamefont {Hu}\ \emph {et~al.}(2024)\citenamefont {Hu},
  \citenamefont {Gu}, \citenamefont {Majumder}, \citenamefont {Ren},
  \citenamefont {Zhang}, \citenamefont {Wang}, \citenamefont {You},
  \citenamefont {Minev}, \citenamefont {Yelin},\ and\ \citenamefont
  {Seif}}]{hu2024demonstration}%
  \BibitemOpen
  \bibfield  {author} {\bibinfo {author} {\bibfnamefont {H.-Y.}\ \bibnamefont
  {Hu}}, \bibinfo {author} {\bibfnamefont {A.}~\bibnamefont {Gu}}, \bibinfo
  {author} {\bibfnamefont {S.}~\bibnamefont {Majumder}}, \bibinfo {author}
  {\bibfnamefont {H.}~\bibnamefont {Ren}}, \bibinfo {author} {\bibfnamefont
  {Y.}~\bibnamefont {Zhang}}, \bibinfo {author} {\bibfnamefont {D.~S.}\
  \bibnamefont {Wang}}, \bibinfo {author} {\bibfnamefont {Y.-Z.}\ \bibnamefont
  {You}}, \bibinfo {author} {\bibfnamefont {Z.}~\bibnamefont {Minev}}, \bibinfo
  {author} {\bibfnamefont {S.~F.}\ \bibnamefont {Yelin}},\ and\ \bibinfo
  {author} {\bibfnamefont {A.}~\bibnamefont {Seif}},\ }\bibfield  {title}
  {\bibinfo {title} {{Demonstration of Robust and Efficient Quantum Property
  Learning with Shallow Shadows}},\ }\href
  {https://doi.org/10.48550/arXiv.2402.17911} {\bibfield  {journal} {\bibinfo
  {journal} {arXiv preprint arXiv:2402.17911}\ } (\bibinfo {year}
  {2024})}\BibitemShut {NoStop}%
\bibitem [{\citenamefont {Hu}\ and\ \citenamefont
  {You}(2022)}]{hu2022hamiltonian}%
  \BibitemOpen
  \bibfield  {author} {\bibinfo {author} {\bibfnamefont {H.-Y.}\ \bibnamefont
  {Hu}}\ and\ \bibinfo {author} {\bibfnamefont {Y.-Z.}\ \bibnamefont {You}},\
  }\bibfield  {title} {\bibinfo {title} {Hamiltonian-driven shadow tomography
  of quantum states},\ }\href
  {https://doi.org/10.1103/PhysRevResearch.4.013054} {\bibfield  {journal}
  {\bibinfo  {journal} {Phys. Rev. Research}\ }\textbf {\bibinfo {volume}
  {4}},\ \bibinfo {pages} {013054} (\bibinfo {year} {2022})}\BibitemShut
  {NoStop}%
\bibitem [{\citenamefont {Cha}\ \emph {et~al.}(2022)\citenamefont {Cha},
  \citenamefont {Ginsparg}, \citenamefont {Wu}, \citenamefont {Carrasquilla},
  \citenamefont {McMahon},\ and\ \citenamefont {Kim}}]{Cha2020}%
  \BibitemOpen
  \bibfield  {author} {\bibinfo {author} {\bibfnamefont {P.}~\bibnamefont
  {Cha}}, \bibinfo {author} {\bibfnamefont {P.}~\bibnamefont {Ginsparg}},
  \bibinfo {author} {\bibfnamefont {F.}~\bibnamefont {Wu}}, \bibinfo {author}
  {\bibfnamefont {J.}~\bibnamefont {Carrasquilla}}, \bibinfo {author}
  {\bibfnamefont {P.~L.}\ \bibnamefont {McMahon}},\ and\ \bibinfo {author}
  {\bibfnamefont {E.-A.}\ \bibnamefont {Kim}},\ }\bibfield  {title} {\bibinfo
  {title} {Attention-based quantum tomography},\ }\href
  {https://doi.org/10.1088/2632-2153/ac362b} {\bibfield  {journal} {\bibinfo
  {journal} {Mach. Learn.: Sci. Technol.}\ }\textbf {\bibinfo {volume} {3}},\
  \bibinfo {pages} {01LT01} (\bibinfo {year} {2022})}\BibitemShut {NoStop}%
\bibitem [{\citenamefont {Hu}\ \emph {et~al.}(2023)\citenamefont {Hu},
  \citenamefont {Choi},\ and\ \citenamefont {You}}]{Hu2023}%
  \BibitemOpen
  \bibfield  {author} {\bibinfo {author} {\bibfnamefont {H.-Y.}\ \bibnamefont
  {Hu}}, \bibinfo {author} {\bibfnamefont {S.}~\bibnamefont {Choi}},\ and\
  \bibinfo {author} {\bibfnamefont {Y.-Z.}\ \bibnamefont {You}},\ }\bibfield
  {title} {\bibinfo {title} {Classical shadow tomography with locally scrambled
  quantum dynamics},\ }\href {https://doi.org/10.1103/PhysRevResearch.5.023027}
  {\bibfield  {journal} {\bibinfo  {journal} {Phys. Rev. Research}\ }\textbf
  {\bibinfo {volume} {5}},\ \bibinfo {pages} {023027} (\bibinfo {year}
  {2023})}\BibitemShut {NoStop}%
\bibitem [{\citenamefont {Torlai}\ and\ \citenamefont
  {Melko}(2018)}]{Torlai2018}%
  \BibitemOpen
  \bibfield  {author} {\bibinfo {author} {\bibfnamefont {G.}~\bibnamefont
  {Torlai}}\ and\ \bibinfo {author} {\bibfnamefont {R.~G.}\ \bibnamefont
  {Melko}},\ }\bibfield  {title} {\bibinfo {title} {{Latent Space Purification
  via Neural Density Operators}},\ }\href
  {https://doi.org/10.1103/PhysRevLett.120.240503} {\bibfield  {journal}
  {\bibinfo  {journal} {Phys. Rev. Lett.}\ }\textbf {\bibinfo {volume} {120}},\
  \bibinfo {pages} {240503} (\bibinfo {year} {2018})}\BibitemShut {NoStop}%
\bibitem [{\citenamefont {Hastings}\ \emph {et~al.}(2010)\citenamefont
  {Hastings}, \citenamefont {Gonz\'alez}, \citenamefont {Kallin},\ and\
  \citenamefont {Melko}}]{hasting2010swap}%
  \BibitemOpen
  \bibfield  {author} {\bibinfo {author} {\bibfnamefont {M.~B.}\ \bibnamefont
  {Hastings}}, \bibinfo {author} {\bibfnamefont {I.}~\bibnamefont
  {Gonz\'alez}}, \bibinfo {author} {\bibfnamefont {A.~B.}\ \bibnamefont
  {Kallin}},\ and\ \bibinfo {author} {\bibfnamefont {R.~G.}\ \bibnamefont
  {Melko}},\ }\bibfield  {title} {\bibinfo {title} {{Measuring Renyi
  Entanglement Entropy in Quantum Monte Carlo Simulations}},\ }\href
  {https://doi.org/10.1103/PhysRevLett.104.157201} {\bibfield  {journal}
  {\bibinfo  {journal} {Phys. Rev. Lett.}\ }\textbf {\bibinfo {volume} {104}},\
  \bibinfo {pages} {157201} (\bibinfo {year} {2010})}\BibitemShut {NoStop}%
\bibitem [{\citenamefont {Kulchytskyy}(2019)}]{kulchytskyy2019renyi}%
  \BibitemOpen
  \bibfield  {author} {\bibinfo {author} {\bibfnamefont {B.}~\bibnamefont
  {Kulchytskyy}},\ }\bibfield  {title} {\bibinfo {title} {{Probing universality
  with entanglement entropy via quantum Monte Carlo}},\ }\href
  {http://hdl.handle.net/10012/15006} {\bibfield  {journal} {\bibinfo
  {journal} {Ph.D. thesis, University of Waterloo}\ } (\bibinfo {year}
  {2019})}\BibitemShut {NoStop}%
\bibitem [{\citenamefont {Beach}\ \emph {et~al.}(2019)\citenamefont {Beach},
  \citenamefont {Vlugt}, \citenamefont {Golubeva}, \citenamefont {Huembeli},
  \citenamefont {Kulchytskyy}, \citenamefont {Luo}, \citenamefont {Melko},
  \citenamefont {Merali},\ and\ \citenamefont {Torlai}}]{qucumber2019}%
  \BibitemOpen
  \bibfield  {author} {\bibinfo {author} {\bibfnamefont {M.~J.~S.}\
  \bibnamefont {Beach}}, \bibinfo {author} {\bibfnamefont {I.~D.}\ \bibnamefont
  {Vlugt}}, \bibinfo {author} {\bibfnamefont {A.}~\bibnamefont {Golubeva}},
  \bibinfo {author} {\bibfnamefont {P.}~\bibnamefont {Huembeli}}, \bibinfo
  {author} {\bibfnamefont {B.}~\bibnamefont {Kulchytskyy}}, \bibinfo {author}
  {\bibfnamefont {X.}~\bibnamefont {Luo}}, \bibinfo {author} {\bibfnamefont
  {R.~G.}\ \bibnamefont {Melko}}, \bibinfo {author} {\bibfnamefont
  {E.}~\bibnamefont {Merali}},\ and\ \bibinfo {author} {\bibfnamefont
  {G.}~\bibnamefont {Torlai}},\ }\bibfield  {title} {\bibinfo {title}
  {{QuCumber: wavefunction reconstruction with neural networks}},\ }\href
  {https://doi.org/10.21468/SciPostPhys.7.1.009} {\bibfield  {journal}
  {\bibinfo  {journal} {SciPost Phys.}\ }\textbf {\bibinfo {volume} {7}},\
  \bibinfo {pages} {009} (\bibinfo {year} {2019})}\BibitemShut {NoStop}%
\bibitem [{\citenamefont {Liu}\ \emph {et~al.}(2023)\citenamefont {Liu},
  \citenamefont {Hao},\ and\ \citenamefont {Hu}}]{liu2023predicting}%
  \BibitemOpen
  \bibfield  {author} {\bibinfo {author} {\bibfnamefont {Z.}~\bibnamefont
  {Liu}}, \bibinfo {author} {\bibfnamefont {Z.}~\bibnamefont {Hao}},\ and\
  \bibinfo {author} {\bibfnamefont {H.-Y.}\ \bibnamefont {Hu}},\ }\bibfield
  {title} {\bibinfo {title} {{Predicting Arbitrary State Properties from Single
  Hamiltonian Quench Dynamics}},\ }\href
  {https://doi.org/10.48550/arXiv.2311.00695} {\bibfield  {journal} {\bibinfo
  {journal} {arXiv preprint arXiv:2311.00695}\ } (\bibinfo {year}
  {2023})}\BibitemShut {NoStop}%
\bibitem [{\citenamefont {Aaronson}\ and\ \citenamefont
  {Gottesman}(2004)}]{aaronson2004improved}%
  \BibitemOpen
  \bibfield  {author} {\bibinfo {author} {\bibfnamefont {S.}~\bibnamefont
  {Aaronson}}\ and\ \bibinfo {author} {\bibfnamefont {D.}~\bibnamefont
  {Gottesman}},\ }\bibfield  {title} {\bibinfo {title} {Improved simulation of
  stabilizer circuits},\ }\href@noop {} {\bibfield  {journal} {\bibinfo
  {journal} {Physical Review A—Atomic, Molecular, and Optical Physics}\
  }\textbf {\bibinfo {volume} {70}},\ \bibinfo {pages} {052328} (\bibinfo
  {year} {2004})}\BibitemShut {NoStop}%
\bibitem [{\citenamefont {Paszke}\ \emph {et~al.}(2019)\citenamefont {Paszke},
  \citenamefont {Gross}, \citenamefont {Massa}, \citenamefont {Lerer},
  \citenamefont {Bradbury}, \citenamefont {Chanan}, \citenamefont {Killeen},
  \citenamefont {Lin}, \citenamefont {Gimelshein}, \citenamefont {Antiga},
  \citenamefont {Desmaison}, \citenamefont {K{\"o}pf}, \citenamefont {Yang},
  \citenamefont {DeVito}, \citenamefont {Raison}, \citenamefont {Tejani},
  \citenamefont {Chilamkurthy}, \citenamefont {Steiner}, \citenamefont {Fang},
  \citenamefont {Bai},\ and\ \citenamefont {Chintala}}]{Paszke2019}%
  \BibitemOpen
  \bibfield  {author} {\bibinfo {author} {\bibfnamefont {A.}~\bibnamefont
  {Paszke}}, \bibinfo {author} {\bibfnamefont {S.}~\bibnamefont {Gross}},
  \bibinfo {author} {\bibfnamefont {F.}~\bibnamefont {Massa}}, \bibinfo
  {author} {\bibfnamefont {A.}~\bibnamefont {Lerer}}, \bibinfo {author}
  {\bibfnamefont {J.}~\bibnamefont {Bradbury}}, \bibinfo {author}
  {\bibfnamefont {G.}~\bibnamefont {Chanan}}, \bibinfo {author} {\bibfnamefont
  {T.}~\bibnamefont {Killeen}}, \bibinfo {author} {\bibfnamefont
  {Z.}~\bibnamefont {Lin}}, \bibinfo {author} {\bibfnamefont {N.}~\bibnamefont
  {Gimelshein}}, \bibinfo {author} {\bibfnamefont {L.}~\bibnamefont {Antiga}},
  \bibinfo {author} {\bibfnamefont {A.}~\bibnamefont {Desmaison}}, \bibinfo
  {author} {\bibfnamefont {A.}~\bibnamefont {K{\"o}pf}}, \bibinfo {author}
  {\bibfnamefont {E.}~\bibnamefont {Yang}}, \bibinfo {author} {\bibfnamefont
  {Z.}~\bibnamefont {DeVito}}, \bibinfo {author} {\bibfnamefont
  {M.}~\bibnamefont {Raison}}, \bibinfo {author} {\bibfnamefont
  {A.}~\bibnamefont {Tejani}}, \bibinfo {author} {\bibfnamefont
  {S.}~\bibnamefont {Chilamkurthy}}, \bibinfo {author} {\bibfnamefont
  {B.}~\bibnamefont {Steiner}}, \bibinfo {author} {\bibfnamefont
  {L.}~\bibnamefont {Fang}}, \bibinfo {author} {\bibfnamefont {J.}~\bibnamefont
  {Bai}},\ and\ \bibinfo {author} {\bibfnamefont {S.}~\bibnamefont
  {Chintala}},\ }\bibfield  {title} {\bibinfo {title} {{PyTorch: An Imperative
  Style, High-Performance Deep Learning Library}},\ }\href
  {http://arxiv.org/abs/1912.01703} {\bibfield  {journal} {\bibinfo  {journal}
  {arXiv preprint arXiv:1912.01703}\ } (\bibinfo {year} {2019})}\BibitemShut
  {NoStop}%
\bibitem [{\citenamefont {Gidney}(2021)}]{Gidney2021}%
  \BibitemOpen
  \bibfield  {author} {\bibinfo {author} {\bibfnamefont {C.}~\bibnamefont
  {Gidney}},\ }\bibfield  {title} {\bibinfo {title} {Stim: a fast stabilizer
  circuit simulator},\ }\href {https://doi.org/10.22331/q-2021-07-06-497}
  {\bibfield  {journal} {\bibinfo  {journal} {Quantum}\ }\textbf {\bibinfo
  {volume} {5}},\ \bibinfo {pages} {497} (\bibinfo {year} {2021})}\BibitemShut
  {NoStop}%
\bibitem [{\citenamefont {Kingma}\ and\ \citenamefont
  {Ba}(2014)}]{kingma2014adam}%
  \BibitemOpen
  \bibfield  {author} {\bibinfo {author} {\bibfnamefont {D.~P.}\ \bibnamefont
  {Kingma}}\ and\ \bibinfo {author} {\bibfnamefont {J.}~\bibnamefont {Ba}},\
  }\bibfield  {title} {\bibinfo {title} {{Adam: A Method for Stochastic
  Optimization}},\ }\href@noop {} {\bibfield  {journal} {\bibinfo  {journal}
  {arXiv preprint arXiv:1412.6980}\ } (\bibinfo {year} {2014})}\BibitemShut
  {NoStop}%
\bibitem [{\citenamefont {Loshchilov}\ and\ \citenamefont
  {Hutter}(2016)}]{Loshchilov2016COSINE}%
  \BibitemOpen
  \bibfield  {author} {\bibinfo {author} {\bibfnamefont {I.}~\bibnamefont
  {Loshchilov}}\ and\ \bibinfo {author} {\bibfnamefont {F.}~\bibnamefont
  {Hutter}},\ }\bibfield  {title} {\bibinfo {title} {{SGDR: Stochastic Gradient
  Descent with Warm Restarts}},\ }\href
  {https://doi.org/10.48550/arXiv.1608.03983} {\bibfield  {journal} {\bibinfo
  {journal} {arXiv preprint arXiv:1608.03983}\ } (\bibinfo {year}
  {2016})}\BibitemShut {NoStop}%
\bibitem [{\citenamefont {Chen}\ and\ \citenamefont
  {Ye}(2011)}]{Chen2011Simplex}%
  \BibitemOpen
  \bibfield  {author} {\bibinfo {author} {\bibfnamefont {Y.}~\bibnamefont
  {Chen}}\ and\ \bibinfo {author} {\bibfnamefont {X.}~\bibnamefont {Ye}},\
  }\bibfield  {title} {\bibinfo {title} {{Projection Onto A Simplex}},\ }\href
  {http://arxiv.org/abs/1101.6081} {\bibfield  {journal} {\bibinfo  {journal}
  {arXiv preprint arXiv:1101.6081}\ } (\bibinfo {year} {2011})}\BibitemShut
  {NoStop}%
\end{thebibliography}%

\end{document}